\documentclass{article}
\usepackage{spconf,amsmath,graphicx}
\usepackage{booktabs}
\usepackage{float}
\usepackage{multirow}
\usepackage{url}

\usepackage{color,soul}

\title{Investigating End-to-End ASR Architectures for Long form Audio Transcription}
\name{\begin{tabular}{c} Nithin Rao Koluguri, Samuel Kriman, Georgy Zelenfroind, Somshubra Majumdar,\\ Dima Rekesh, Vahid Noroozi, Jagadeesh Balam, Boris Ginsburg
\end{tabular}
}
\address{NVIDIA}
\begin{document}
%\ninept
%
\maketitle
\begin{abstract}

 This paper presents an overview and evaluation of some of the end-to-end ASR models on long-form audios. We study three categories of Automatic Speech Recognition(ASR) models based on their core architecture: (1) convolutional, (2) convolutional with squeeze-and-excitation and (3) convolutional models with attention. 
We selected one ASR model from each category and evaluated Word Error Rate, maximum audio length and real-time factor for each model on a variety of long audio benchmarks: Earnings-21 and 22, CORAAL, and TED-LIUM3. The model from the category of self-attention with local attention and global token has the best accuracy comparing to other architectures. We also compared models with CTC and RNNT decoders and showed that CTC-based models are more robust and efficient than RNNT on long form audio.
% 
% talk about some of the interesting findings
\end{abstract}
\begin{keywords}
Automatic Speech Recognition (ASR), Long-form Audio, Earnings-21, CORAAL, TED-LIUM
\end{keywords}
\section{Introduction}
\label{sec:intro}

Long-form speech presents unique challenges for automatic speech recognition (ASR). While there is no strict time limit that defines ``long-form'', it generally refers to audio recordings that can range from several minutes to several hours. Long-form audio is often encountered in various applications, such as transcription services, podcasting, audio book production, and more.
End-to-end ASR models are usually trained on short speech utterances of up to 30 seconds in length. 
Most of the common benchmarks  used in ASR research are also short-form,  so some of popular models may not be able to transcribe on long-form audio.
 For example, the state-of-the-art Conformer Large model \cite{gulati20_interspeech} can only handle audio up to 12 minutes long on a A6000 GPU with 48 GB memory. The maximum utterance length during inference depends on the model architecture, and it is  mainly limited by the device memory.

One way to overcome memory limitations during long-form audio inference is to use streaming ASR methods, for example split the input split into smaller chunks. 
ASR outputs of individual chunks are then merged to get the final transcription. Another method is to use models specially designed for streaming.  An implementation of such a model in NeMo \cite{Harper_NeMo_a_toolkit} converts the Conformer's non-autoregressive encoder into an autoregressive recurrent model during inference using a cache for activations computed from previous timesteps. This work mainly investigates if we can use end-to-end ASR models that are trained on short-form audio to transcribe long-form audio. 
There are a number of ways to transcribe long-form audio using end-to-end ASR models trained on short utterances. For example, one can use a voice activity detector (VAD)  to segment the audio at long pauses or silences and then transcribe each segment independently. Other methods \cite{huang2023e2e, huang2022e2e} improve transcription accuracy compared to using VAD by  predicting segmentation labels. Another approach is to split the incoming audio into overlapped chunks and then merge the ASR outputs of each chunk.  For example, the authors of \cite{chiu2019comparison},  present an overlappping inference for attention-based models where long audio is broken into fixed length overlapped segments,  and a matching algorithm is used to merge the results to reduce errors at segment boundaries. A similar method is available for both CTC and RNN-T models in NeMo \cite{nemo_buffered_asr}.

This paper is mainly focused on the single pass offline transcription of long-form audio. We conduct a comprehensive evaluation of three primary types of end-to-end ASR models for long audio: 
\begin{itemize}
    \item QuartzNet \cite{kriman2020quartznet} model based on depth-wise separable convolution 
    \item ContextNet \cite{han2020contextnet} and Citrinet \cite{majumdar2021citrinet} models add "Squeeze-and-Excitation" based global context to convolutions  
    \item Fast Conformer \cite{FastConformer} is a redesigned for long audio Conformer \cite{gulati20_interspeech} with local attention and global tokens
\end{itemize}
Following are the main contributions of this paper:
\begin{enumerate}

  \item We evaluated 3 types of models on ``long speech" benchmarks: Earnings-21 and -22, CORAAL, and TED-LIUM3. For each model we measured: the maximum sequence length which can transcribed in one pass, Word Error Rate (WER), and Real-Time Factor (RTF). 
  \item We investigate the effect of global context on the accuracy of long-form audio transcription.
  \item Finally, we compared the transcription accuracy and efficiency of models with CTC and RNNT decoders on long-form audio transcription.
\end{enumerate}
All models used in the paper, training and inference scripts are open-sourced in NeMo toolkit. 

\begin{figure*}[t]
    \centering
    \includegraphics[width=0.9\textwidth]{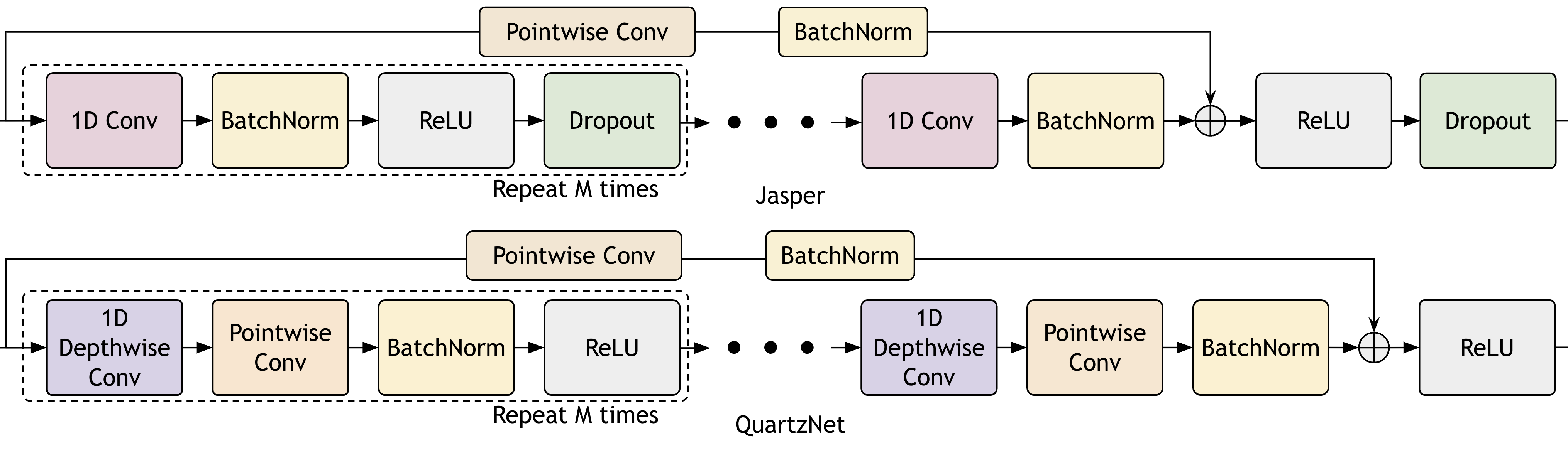}
    \caption{Jasper and QuartzNet block comparison:  QuartzNet replaces 1D convolution with 1D depthwise-separable convolution, consisting of a depthwise and pointwise layers}
    \label{fig:conv_model_arch}
\end{figure*}

\section{Related Work}
There are a number of ways to transcribe long-form audio using end-to-end ASR models trained on short utterances. For example, one can use a voice activity detector (VAD)  to segment the audio at long pauses or silences and then transcribe each segment independently. Other methods \cite{huang2023e2e, huang2022e2e} improve transcription accuracy compared to using VAD by  predicting segmentation labels. Another approach is to split the incoming audio into overlapped chunks and then merge the ASR outputs of each chunk.  For example, the authors of \cite{chiu2019comparison},  present an overlappping inference for attention-based models where long audio is broken into fixed length overlapped segments,  and a matching algorithm is used to merge the results to reduce errors at segment boundaries. A similar method is available for both CTC and RNN-T models in NeMo \cite{nemo_buffered_asr}. The merging process may introduce errors  especially when the ASR outputs are not well aligned with the audio. See \cite{chiu2019comparison} for more detailed overview of  method for decoding long audios based on segmentation.

Another method to transcribe long audio is based on streaming ASR models \cite{naray_longform,li2021better,yu2020dual,yao2021wenet}. For example, a streaming Conformer in NeMo converts non-autoregressive encoder into an autoregressive recurrent model during inference. This  drastically reduces the computation cost when compared to traditional buffer-based methods by using a cache to store the activations. The stored in cache intermediate activations are used in future steps. The model with activation cache does not need any buffer or overlapping chunk, so there are no unnecessary duplicated computations.  The model has also limited right and left contexts during training to maintain consistent conditions during training and streaming inference. Note that the model is still trained efficiently in non-autoregressive mode, similar to offline models. There are also other advanced methods \cite{chang_contextaware,hori2021advanced} that operate on segmented long audios but the context from previously decoded utterances is propagated as context for decoding subsequent utterances.

In  this paper,  we only look only at methods for "one pass" offline transcription of long audios without additional segmentation. We believe handling long audios in one shot as single-pass inference has the following advantages:
\begin{enumerate}
 
 \item Enables the inclusion of complete acoustic context during decoding.
 \item Eliminates the need for post-processing to merge the hypothesis from individual chunks.
 \item Allows for the application of continuous beam search algorithms due to the absence of merging steps.
 % \item Enables future research into end-to-end training for long-form audio.
\end{enumerate}

\section{ASR Models for Long audio}
\label{sec:models}

\begin{figure*}[t]
    \centering
    \includegraphics[width=0.9\textwidth]{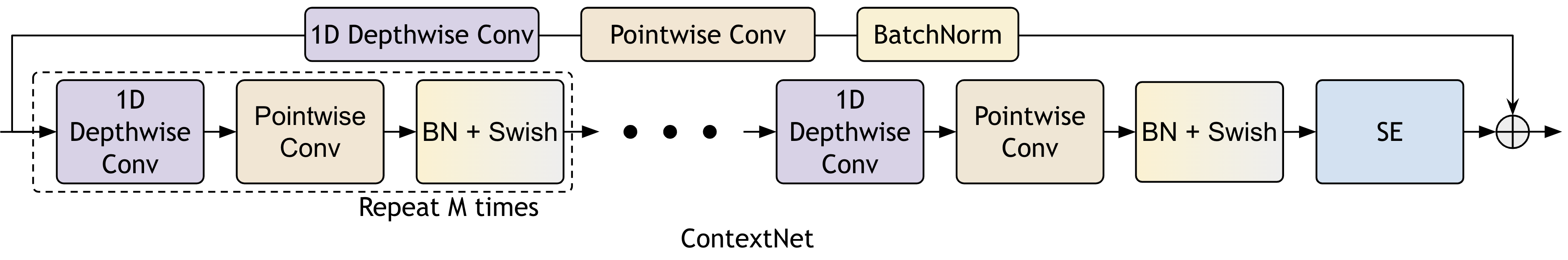}
    \caption{ContextNet adds a squeeze-and-excitation module (SE) and the end of the block to incorporate global information}
    \label{fig:enter-label}
\end{figure*}

We selected three models for long audio ASR: QuartzNet2 -- convolution-only model,  ContextNet -- the convolution + SE-based global context model, and the Fast Conformer -- model with local attention and global tokens. 

\subsection{Convolution-only based models}

Convolutional neural networks (CNNs) are well suited to capturing local temporal patterns in audio, making them a natural choice for ASR. One of the first  convolutional ASR models was 
Wav2Letter \cite{collobert2016wav2letter}. By using strided 1D convolutions near the initial layers with raw waveform and power-spectrum features, Wav2Letter managed to speed up the most computationally intensive parts of the network, achieving impressive efficiency. 
The Jasper model \cite{li19_interspeech}, added residual connections to Wav2Letter, which allow to increase depth of model to  54 layers. 
Jasper consists of a series of  blocks, where each block applies a sequence of operations: 1D-convolution, batch normalization (BN), ReLU, and dropout (see Fig.\ref{fig:conv_model_arch}). Residual connections link the input and output of each block. 

QuartzNet \cite{kriman2020quartznet} improved Jasper  by replacing 1D convolution layers with 1D time-channel separable convolution  (Fig.\ref{fig:conv_model_arch}). 1D time-channel separable convolution block consists of a 1D depthwise convolution layer with kernel length $K$ that operates on each channel individually but across $K$ time frames, and a pointwise convolution layer that operates on each time frame independently but across all channels. 1D time-channel separable convolution can operate in a similar way to standard convolution, while having significantly less parameters: QuartzNet  with 22M parameters achieves the accuracy similar to the Jasper with 333M parameters.

In this study we use QuartzNet 2.0 - an updated and scaled-up version of the original QuartzNet. In order to improve QuartzNet, we introduce several modifications. Firstly, we unify the 1D depthwise convolutional layers by setting all of their kernel sizes to 7. By reducing the kernel sizes, we can achieve better streaming performance. In addition, we add another downsampling layer with stride 2 to the beginning of the encoder, which doubles the overall downsampling rate from 2x to 4x for increased efficiency. Unlike the original, QuartzNet 2.0 is also trained with a hybrid CTC-RNNT decoder, and uses word-piece tokenization. Hybrid CTC-RNNT ASR models are trained with two decoders of CTC and RNNT in a jointly manner. It would enable use to just train one model instead of two separate models. It also reduces the number of steps needed for the convergence of the CTC model with the help from the RNNT decoder.

\subsection{Convolutional model with  Squeeze-and-Excitation global context}

ContextNet \cite{han2020contextnet} is  convolutional RNN-Transducer \cite{graves2012rnnt}  module that is enhanced with 1D Squeeze-and-Excitation (SE) \cite{han2020contextnet} global context modules. Like the QuartzNet, it utilizes 1D time-channel separable convolutions. Deviating from the original QuartzNet, it uses the same convolution kernel size of 5 throughout the model and utilizes the SiLU (Swish) activation  \cite{ramachandran2017swish}. ContextNet replaces the CTC decoder with a Transducer decoder.  
% and attains superior recognition accuracy on standard short-form ASR benchmarks comparing to CTC models without an external language model. 
The ContextNet starts with a prolog block, followed by 22 blocks, grouped together into 4 segments. Each module in a given segment shares the same number of output features, scaled by $\alpha$ in order to increase or decrease the size of the model, but at the beginning of each subsequent segment, the number of output features is doubled.  The first three segments end with a 1D time-channel separable convolutional layer with stride 2, so ContextNet progressively down-samples the input three times in the time domain, and has an output resolution of $80$ms.  

Citrinet \cite{majumdar2021citrinet} is a ContextNet-like model with encoder which was  modified to use a CTC decoder. 
 
Like ContextNet, Citrinet uses a standard acoustic front-end: 80-dimensional log-mel filter banks with a $25$ms window and a stride of $10$ms and performs progressive downsampling in the first three segments, thereby having an output resolution of $80$ms. 
Deviating from ContextNet, it performs the downsampling at the beginning of each of the segments rather than at the end. In addition, all blocks throughout the entire network  share the same input and output dimension. 
% (controlled by $C$). 
Finally, unlike the uniform convolution kernel size utilized in ContextNet, Citrinet designs a specific layout of kernels across each of its blocks that was found to bring more stable and accurate results when utilizing a CTC decoder.

\begin{figure*}[t]
    \centering
    \includegraphics[width=0.9\textwidth]{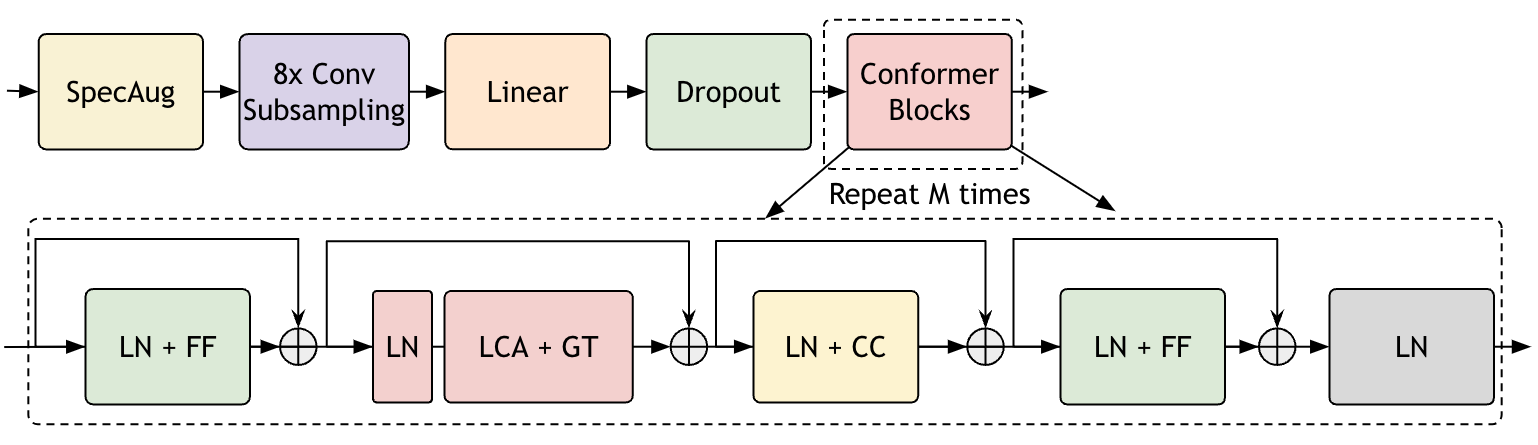}
    \caption{Fast Conformer. Input sequence is sub-sampled at an `8x' rate and processed through modified Conformer blocks. Each block contains FF (Feed Forward), Multi Head  Attention, and CC (Conformer Convolution) modules, separated by Layer normalization (LN). Fast Conformer uses Limited Context Attention and Global Token (LCA + GT) instead of regular Mulit-head Attention (MHA) used in the original Conformer .}
    \label{fig:conformer}
\end{figure*}

\subsection{Convolution + Attention Based Models}

Building upon the previously mentioned models, a newer class integrates both convolutions and attention mechanism for speech recognition. These architectures aim to blend the localized pattern recognition of convolutional structures with the global contextual representations created by attention mechanisms. Among the representatives in this category, the Conformer has become a particularly influential model.

%\subsubsection{Conformer}

The Conformer \cite{gulati20_interspeech} architecture incorporates elements of both convolutional neural networks and Transformers. Its design consists of modular blocks, each encompassing feed-forward networks, convolutional modules, and multi-head self attention. Conformer-RNNT, the variant of this model with a transducer decoder, obtains state of the art results on various speech benchmarks. However, the quadratic time and memory complexity of attention with respect to sequence length makes it more compute-heavy than convolutional model for long audio, and significantly limits the maximum duration that can be processed with this model.

\begin{figure}[t]
    \centering
    \includegraphics[width=0.45\textwidth]{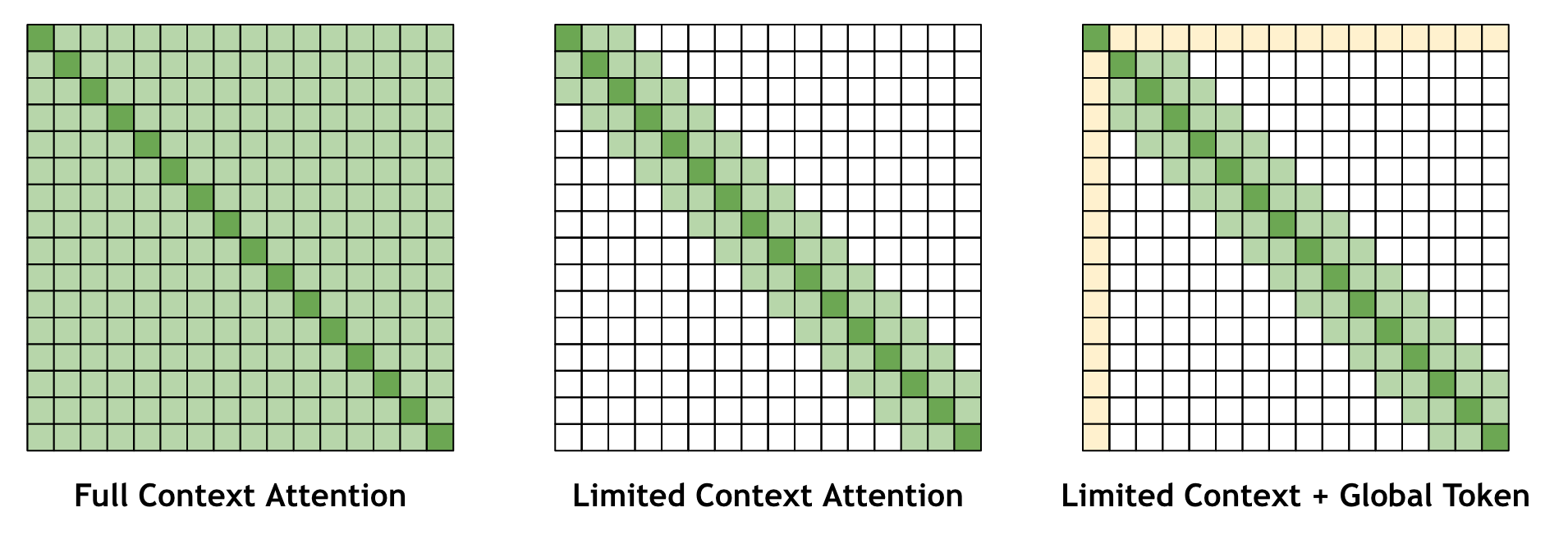}
    \caption{The Fast Conformer model combines local attention with a single global attention token. The left figure depicts full-context attention, revealing the global context from self-attention modules. The middle figure illustrates Limited Context Attention (LCA), while the right figure demonstrates the incorporation of the global token (GT) with limited context attention for global context.}
    \label{fig:fc_attn}
\end{figure}

Fast Conformer \cite{FastConformer} (Fig.\ref{fig:conformer}) is a re-designed version of Conformer, optimized for fast inference and more stable scaling while retaining transcription quality. In order to address these challenges, the authors change Fast Conformer's downsampling schema, which accounted for 20\% of the computation time for each forward pass of the Conformer-Large model. The 2D convolutional layers in the downsampling block are changed to depthwise separable convolution, significantly reducing the computation time. An additional 2x downsampling layer is also added, increasing the models overall downsampling rate from 4x to 8x. In each downsampling layer, the number of channels is set to 256 and kernel size is set to 9.

In addition, in order to increase the efficiency of processing long-form audio, the attention layers in Fast Conformer can be replaced with limited context attention (LCA). In this variant of attention, each time step only attends to a limited number of time steps to the left and right side of it, in a sliding window pattern. The size of context on each size is set by default to 128 steps, corresponding to around 10 seconds of audio before downsampling. This attention can be implemented efficiently using the overlapping chunk approach, thus solving the issue of quadratic complexity of attention with respect to audio length, and allowing the model to process much longer audio. Furthermore, by adding a single global token (GT), which can attend and is attended to by all other tokens, the model incorporates global context. By using limited context attention in combination with a single global attention token (LCA + GT) Fig. \ref{fig:fc_attn}, Fast Conformer can be used to efficiently transcribe long audio up to 11 hours on A100 and up to 8 hrs on A6000 in a single forward pass with good results.

\subsection{Training}
\label{sec:datasets}

% \subsection{Training Datasets}
All models were trained on the same 25,000 hours of public  speech data combined from LibriSpeech (LS) \cite{panayotov2015librispeech}, the English part of Multilingual LibriSpeech (MLS) \cite{MLS}, Mozilla Common Voice \cite{Mozilla}, Wall Street Journal (WSJ) \cite{wsj}, Fisher \cite{cieri2004fisher}, Switchboard-1 \cite{mihatschgodfrey}, National Speech Corpus (NSC) \cite{koh2019building}, Voxpopuli-English subset \cite{wang-etal-2021-voxpopuli}, VCTK \cite{yamagishi2019cstr}, Europal-ASR \cite{europarlasr2021}, and People's Speech \cite{galvez2021people}. 

We used short utterances with maximum duration of 20 sec for training. Each model is trained for 300K steps with a warm-up  of 25K steps. QuartzNet2 and ContextNet were trained using a cosine scheduler and AdamW optimizer. Fast Conformer was trained using a Noam scheduler and AdamW. Fast Conformer models were trained with full attention, and then fine-tuned (FT) with limited context attention (LCA), whether with or without a global token (GT), for an additional 10K steps.

\begin{table*}[t]
\label{tab:datasets}
\caption{Long-form audio evaluation datasets. The audio durations of the datasets vary from 1 minute to over 2 hours.}
\centering
\begin{tabular}{l|c|c|c|c}
\toprule
\textbf{Dataset} & $\begin{array}{c}\textbf{ Number of } \\ \textbf{ Recordings }\end{array}$ & $\begin{array}{c}\textbf { Min duration } \\
(\textbf{min})\end{array}$ & $\begin{array}{c}\textbf { Max Duration } \\
(\textbf{min})\end{array}$ & $\begin{array}{c}\textbf { Mean duration } \\
(\textbf{min})\end{array}$ \\
\midrule
CORAAL & 231 & 0.98 & 81.86 & 35.27 \\
Earnings-21 & 44 & 18.29 & 95.68 & 53.54 \\
Earnings-22 & 125 & 14.58 & 123.45 & 57.55 \\
TED-LIUM 3 & 11 & 6.89 & 29.53 & 16.74 \\
\bottomrule
\end{tabular}
\end{table*}

\section{Long audio Evaluation}
We evaluate all ASR  models for single pass offline inference on long-form audio. All evaluations are conducted using a  A6000 GPU (48GB) with bfloat16 precision and a batch  of 1.

\subsection{Evaluation Datasets}
We evaluate all  models on four English datasets, namely TED-LIUM3 \cite{hernandez2018ted}, Earnings-21 \cite{del2021earnings}, Earnings-22 \cite{speech-datasets},  and CORALL \cite{Farrington_Kendall_2021}, which  contain diverse data of  various lengths and recording conditions.

We used  the test set from TED-LIUM \cite{hernandez2018ted} v.3  which  comprises 11 TED talks, each with an average duration of approximately 16 minutes. 
We sliced the audio files for evaluation based on the onset of the first labeled segment and the end of the final labeled segment of each talk \cite{radford2023robust}.

Earnings-21 \cite{del2021earnings} and Earnings-22 \cite{speech-datasets} are corpora of earnings calls  from  different financial sectors.  
% Both dataset samples have a mixed sampling rate from 11kHz to 44kHz. 
Earnings-21 consists of 39 hours of audio, while the Earnings-22 dataset consists of 119 hours of audio. Both datasets are used to benchmark ASR systems on long-form audio transcription.
Earnings-21 and Earnings-22 contain various entity names and numerical forms, so we applied a normalization process to both the predicted and ground truth texts to all datasets using the Whisper normalizer \cite{radford2023robust}. 
% This normalization ensures a fair comparison among models that have been trained with various tokenizers.

% \subsubsection{CORAAL}
The CORAAL \cite{Farrington_Kendall_2021} (Corpus of Regional African American Language) consists of 231 English language interview recordings, typically involving two-way conversations. The CORAAL dataset consists of strongly accented speech collected during interviews with individuals from diverse age groups, including substantial overlapped speech.
We processed the provided transcripts to remove non-spoken words such as pauses and special characters. All recordings were initially sampled at various sampling rates ranaging from 11kHz to 44.1kHz, and we resampled them to 16kHz. Dataset characteristics of these sets are provided in Table.\ref{tab:datasets}.

\begin{table}[t]
\caption{Maximum audio length for single-pass inference on an A6000 GPU. }
\centering
\scalebox{0.8}
{
    \begin{tabular}{c|c|l|c}
     \toprule
       \textbf{Model} & $\begin{array}{l}\textbf { Size} \\
       \textbf { (M) }\end{array}$ & { } \textbf{Encoder type} & $\begin{array}{c}\textbf{Max Length} \\
       \textbf {(min)}\end{array}$ \\
       \midrule
             % Jasper & 330 & 1D Depth-wise Conv & $\mathrm{CTC}$ & char & 7000 & 633 \\
             % QuartzNet & 20 & Conv & CTC & char & 7000 & 817 \\
             {QuartzNet2} & 120 & 1D depth-wise Conv & 817 \\
             % \midrule
             % CitriNet & 140 & Conv $+$ Context & CTC & bpe & 7000 & 342 \\
             {ContextNet} & 140 &   { } + SE Context  & 342 \\
             % \midrule
             Conformer & 120 &     { } + Attention & 12 \\
             % Fast-Conformer & 114 & Attention & CTC/RNNT & bpe & 25000 & 23 \\
             % $\begin{array}{l}\text{Fast-Conformer} \text { w/ local } \\
            % \text { Attention }\end{array}$ & 114 & Attention & CTC/RNNT & bpe & 25000 & 467 \\
            {Fast Conformer } & 114 &  { } + Local Attention & 467 \\
        \bottomrule
        \end{tabular}
        }
    \label{tab:model_summary}
\end{table}

\subsection{Maximum audio duration}
\label{sec:models_list}

Table \ref{tab:model_summary} provides the maximum audio length for single-pass inference on an A6000 GPU for each model. 
Convolution-only  models, such as QuartzNet2, can  process audio for durations exceeding 12 hours. ContextNet also shows very good long audio  capabilities, exceeding 5 hours of offline audio processing. In contrast, attention-based Conformer can handle audio sequences up to a maximum of 12 minutes.
Recognizing the need for extended audio processing capabilities, the redesigned Fast Conformer, incorporating localized attention mechanisms,  can transcribe audio sequences spanning up to a substantial 8 hour duration, reducing the gap between convolutional-only  and attention-based models.

\subsection{Real-Time Factor (RTF) }
\begin{figure}[t]
 \label{fig:rtf-graph}
 \centering
 \includegraphics[width=0.48\textwidth]{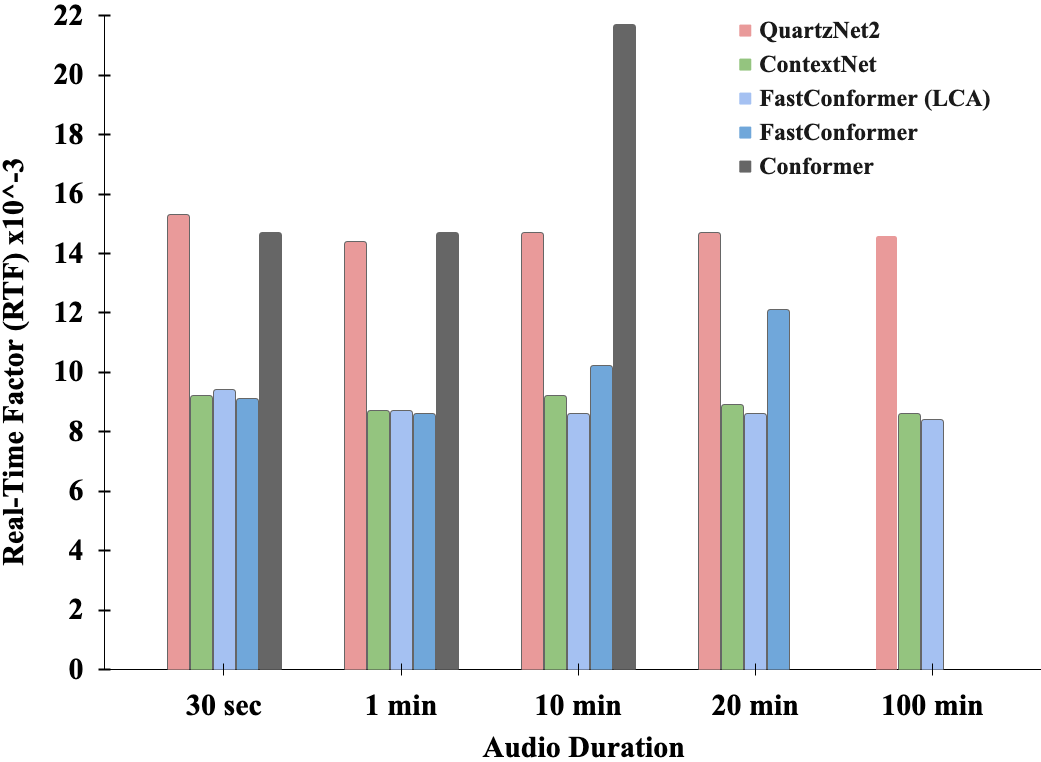}
    \caption{Real-Time Factor (RTF) vs audio duration for QuartzNet2, ContextNet, Conformer, Fast Conformer with full and limited context attention (LCA). All models have been evaluated with RNNT decoder at various durations.  Lower values indicate better performance. On average, the Fast Conformer with limited context attention outperforms the Convolution and ContextNet-based models. The maximum audio processing limit of Conformer with full attention on A6000 is 12 minutes, while Fast Conformer with full attention can process up to 23 minutes.}
    % RTF is computed as ratio of time required to transcribe audio to audio length.
\end{figure}

The Real-Time Factor (RTF) metric is used to quantify the efficiency of these models in processing long audio samples: 

\begin{equation*}
RTF = \frac{\text{Time to transcribe the Audio}}{\text{Audio Duration}}
\end{equation*}
A lower RTF indicates that the model can transcribe long audio samples faster. 
We measure RTF at various durations to evaluate the inference speed of the models. The RTF scores for all models except Conformer, consistently remain within specific ranges at varying audio durations, demonstrating they decode audio in a duration length-agnostic manner. 

In Fig. \ref{fig:rtf-graph} we present RTF for  QuartzNet2, ContextNet, Conformer, and Fast Conformer models with an RNNT decoder across various audio durations. Notably, the Fast Conformer with limited context attention (LCA) exhibits superior efficiency compared to the other models, evidenced by its decreasing RTF with longer audio durations. This improvement can be attributed to the Fast Conformer's 8$\times$ subsampling, in contrast to the 4$\times$ subsampling used in convolution-only-based models. For comparison, we also include an RTF plot for Conformer with full attention.  Although Conformer model demonstrates exceptional accuracy on short audio benchmarks \cite{gulati20_interspeech}, their maximum duration is very short. For example, for Conformer-Large is limited to 12-minute on an A6000 GPU with 48GB of RAM.  Overall, the Fast Conformer model stands out as an efficient attention-based model for processing long-form audio. 

\subsection{Accuracy}

% \subsection{Effect of Global Context}
\begin{table}[t]
\label{tab:context}
 \centering
 \caption{The effect of global context on model accuracy. We compare  RNNT models: QuartzNet2, ContextNet, and three variants of Fast Conformer with Limited Context Attention (LCA): (1) No fine-tuning (2)  Fine-tuned, (3) Fine-tuned with LCA and global token (FT+LCA+GT). Greedy WER (\%). }
 \scalebox{0.70}
 {
  \begin{tabular}{l|c c c c}
   \toprule
   \textbf{Model} & \textbf{TED-LIUM3} & \textbf{Earnings21} & \textbf{Earnings22} & \textbf{CORAAL} \\
   \midrule
   QuartzNet2 & 7.31 & 23.1 & 31.17 & 40.64 \\
   ContextNet & 5.52 & 19.12 & 24.37 & 38.75 \\
  % \midrule
   Fast Conformer (LCA) & 5.88 & 17.08 & 24.67 & 37.35 \\
   % $\begin{array}{c}\text { + Loca } \\
   % \text{ Finetuning }\end{array}$ & 5.08 & 14.82 & 20.44 & 30.28 \\
   { } + FT  & 5.08 & 14.82 & 20.44 & 30.28 \\
   { } { } + GT & 4.98 & 13.84 & 19.49 & 28.75 \\
   \bottomrule       
    \end{tabular}
    }
\end{table}

Global context can significantly enhance the accuracy in long-form audio, which be achieved by integrating global context from audio through neural layers or embeddings.

For evaluation of global context impact on accuracy, we use three type of models with RNNT decoder: QuartzNet2, which lacks global context integration; ContextNet, which incorporates global context via squeeze-and-excitation (SE) modules; and Fast Conformer with local context of (128, 128), which utilizes global token. We use three variants of Fast Conformer with Limited Context Attention (LCA):
\begin{itemize}
    \item LCA: no fine-tuning, no global token.
    \item FT+LCA: finetuned , no global token 
    \item FT+LCA+GT: finetuned with LCA and global tokens 
\end{itemize}
Table. 4
% \ref{tab:context} 
presents the accuracy of QuartzNet2, ContextNet, and Fast Conformer RNNT models on four different datasets with long audios: TED-LIUM3, Earnings-21, Earnings-22, and CORAAL.  

QuartzNet2 that lack global context performs relatively poorly on  long-form audio. The performance of ContextNet dramatically improves on all datasets compared to QuartzNet2, demonstrating the benefit of the SE module on long-form audio. As we transition from convolution-based to attention-based models, the improvement from Fast Conformer with a self-attention layer but limited context does not show significant gains on long audio. However, fine-tuning Fast Conformer with local attention leads to additional improvements. The best WER is achieved when finetuning the Fast Conformer with limited context attention and with a global token that captures global context.

\section{RNNT vs CTC on Long Audio}

\begin{figure}[t]
\label{fig:ctc_rnnt_rtf_graph}
 \centering
 \includegraphics[width=0.48\textwidth]{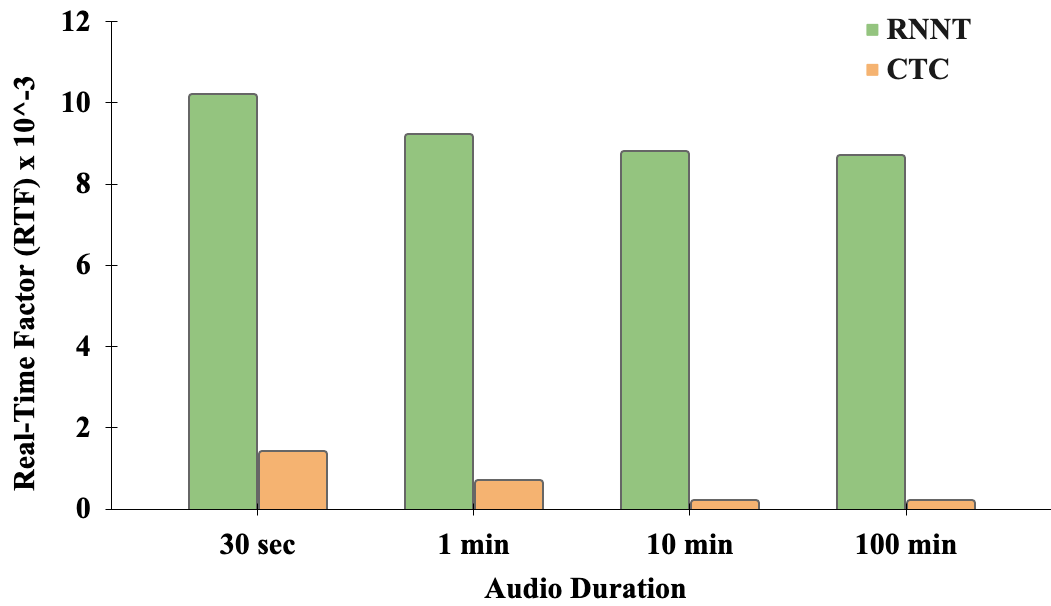}
 \caption{Real-Time Factor (RTF) of Fast Conformer with limited context attention (LCA) model with CTC and RNNT decoder at various audio durations. Lower the better.  RNNT decoder models are on average 43x slower than CTC models for a 100 minute duration audio.}
\end{figure}

ASR models with the Recurrent Neural Network Transducer (RNNT) decoder \cite{rao2017exploring} tends to exhibit higher memory requirements and slower processing speeds when compared to models trained with Connectionist Temporal Classification (CTC) loss \cite{graves2006connectionist}, primarily owing to its intricate RNN structure. While this performance discrepancy may be tolerable for short audio segments during inference, it becomes a critical concern for processing long-form audio. As illustrated in Fig. \ref{fig:ctc_rnnt_rtf_graph}, the RNNT models' RTF is considerably slower, approximately 10x, than the CTC model even for a 30-second audio segment. This discrepancy between RNNT and CTC escalates significantly with audio length, reaching approximately 43x when processing a 100 minute audio utterance, making RNNT less efficient for decoding long-form audio.

The efficiency gap between CTC and RNNT models may be less significant under poor CTC performance. While RNNT decoder-based models outperform CTC models in various short-form audio benchmarks \cite{FastConformer}, it is crucial to assess their validity for long-form audio. To investigate, we compared the performance of QuartzNet2 and Fast Conformer models with limited context (LCA) of (128,128) using CTC and RNNT decoders across all long-form benchmarking datasets (see Table \ref{tab:ctc_rnnt}). The results reveal that the QuartzNet2 model with a CTC decoder outperforms the RNNT decoder across all datasets. Furthermore, the Fast Conformer model with limited context attention (LCA) trained with CTC loss performs equally well compared to the RNNT decoder. However, finetuning the Fast Conformer with limited context attention and global token (FT+LCA+GT) demonstrates that RNNT models perform significantly better than CTC finetuned models, highlighting the efficiency and robustness of CTC models within limited context attention and RNNT models when using global context.

\begin{table}[t]
\label{tab:ctc_rnnt}
\caption{Comparison of QuartzNet2 and Fast Conformer with CTC and RNNT decoders on long-form speech benchmarks. Greedy WER(\%).}
\centering
\scalebox{.65}{
\begin{tabular}{c|c|c|c|c|c}
    \toprule
    \textbf{Model} & \textbf{Decoding} & \textbf{TED-LIUM3} & \textbf{Earnings21} & \textbf{Earnings22} & \textbf{CORAAL} \\
    \midrule
    \multirow{2}{*}{QuartzNet2} & CTC & 6.67 & 19.52 & 26.81 & 40.19 \\
    % \cline{2-6}
    & RNNT & 7.31 & 23.1 & 31.17 & 40.64 \\
    \midrule
    \multirow{2}{*}{\begin{tabular}{@{}c@{}}{Fast Conformer} \\{(LCA)}\end{tabular}} & CTC & 5.64 & 16.86 & 24.24 & 37.79 \\
    % \cline{2-6}
    & RNNT & 5.88 & 17.08 & 24.67 & 37.35 \\
    \midrule
    \multirow{2}{*}{\begin{tabular}{@{}c@{}}{Fast Conformer} \\{(FT+LCA+GT)}\end{tabular}} & CTC & 5.53 & 15.61 & 22.37 & 35.23 \\
    % \cline{2-6}
    & RNNT & 4.98 & 13.84 & 19.49 & 28.75 \\
    % \midrule
    % \multirow{2}{*}{\begin{tabular}{@{}c@{}}Fast Conformer \\ ( + global tokens)\end{tabular}} & \multirow{2}{*}{ RNNT} &  \multirow{2}{*}{4.72} & \multirow{2}{*}{13.11} & \multirow{2}{*}{21.61} & \multirow{2}{*}{25.66} \\
    % &   &     &   &   &   \\
    \bottomrule
    \end{tabular}
    }
\end{table}

To compare CTC and RNNT performance on varying audio segment durations, we conducted additional evaluations using the TED-LIUM3 dataset, which provides timestamps for individual segments. These segments were derived from speaker speech, excluding non-speech segments from the ground truth STM files.  We utilized these segments to create a shorter evaluation set called ``short-form". The audio segments in this set range from 0.35 seconds to 32 seconds, with an average duration of 8.15 seconds. We evaluated both QuartzNet2 and Fast Conformer models on both short and long utterances from the TEDLIUM dataset. Our findings, as presented in Table \ref{tab:short_form2}, reveal that for QuartzNet2 and Fast Conformer architectures with LCA prior to finetuning (LCA), CTC models outperform RNNT models on long-form audio, whereas RNNT models exhibit superior performance when finetuned with global tokens (FT+LCA+GT) on both long and short form utterances. However, the  ``Change in WER" column demonstrates that CTC decoder-based models display greater robustness across a range of audio durations compared to RNNT decoders.

\section{Conclusion}
\label{sec:conclusion}

\begin{table}[t]
\caption{CTC and RNNT decoders for QuartzNet2 and two Fast Conformer-LCA variants: before fine-tuning (LCA) and after fine-tuning with global tokens (FT+LCA+GT). Evaluation  on TED-LIUM3 short-form and long-form audio. The ``Change in WER(\%)" column highlights the robustness of the CTC comparing RNNT when transitioning from long-form to short-form audio. }
\scalebox{0.75}{
\centering
    \begin{tabular}{c|c|c|c|c}
    \toprule
    \textbf{Model} & \textbf{Decoder} & \textbf{Long-form} & \textbf{Short-form} & \textbf{Change in WER} \\
    \midrule
    \multirow{2}{*}{ QuartzNet2 } & CTC  & 6.67 & 6.57 & 0.1 \\
     & RNNT & 7.31 & 6.5 & 0.81  \\
     \midrule
    \multirow{2}{*}{\begin{tabular}{@{}c@{}}Fast Conformer \\ (LCA) \end{tabular}} & CTC & 5.64 & 5.01 & 0.63 \\
     & RNNT & 5.88 & 4.42 & 1.46 \\
     \midrule
    \multirow{2}{*}{\begin{tabular}{@{}c@{}}Fast Conformer \\ (FT+LCA+GT) \end{tabular}} & CTC & 5.53 & 4.89  & 0.64 \\
     & RNNT &  4.98 & 3.97  & 1.01 \\
     % \multirow{2}{*}{\begin{tabular}{@{}c@{}}Fast Conformer \\ (w/ Global Token)\end{tabular}} & \multirow{2}{*}{RNNT} & \multirow{2}{*}{4.02} & \multirow{2}{*}{4.72} \\
     % &  &  &  \\
     \bottomrule
    \end{tabular}
    }
    \label{tab:short_form2}
\end{table}

In this paper, we studied three ASR models: QuartzNet, ContextNet and Fast Conformer on single-pass offline inference task. We evaluated these models using long-form datasets: Earnings-21, Earnings-22, CORAAL, and TED-LIUM v3. For each model we compute WER, RTF, and maximum sequence length which model can transcribe in one shot.  We confirmed the importance of global context within the model for both short and long-form audio transcription. The Fast Conformer model with local attention and global token has best accuracy on  long-form audio. We also demonstrated that models with CTC decoder  are significantly more efficient and robust for long-form audio transcription than RNNT.

% \clearpage

% References should be produced using the bibtex program from suitable
% BiBTeX files (here: strings, refs, manuals). The IEEEbib.bst bibliography
% style file from IEEE produces unsorted bibliography list.
% -------------------------------------------------------------------------
\bibliographystyle{IEEEbib}
\bibliography{strings,refs}

\end{document}